\documentclass[conference]{IEEEtran}

\ifCLASSINFOpdf
\usepackage[pdftex]{graphicx}

\else
\fi
\usepackage{ifpdf}
\usepackage{cite}
\usepackage{subcaption,graphicx,color}
\usepackage{amsmath,amssymb,amsfonts,mathrsfs,amsthm}
\usepackage{algorithmic}
\usepackage[english]{babel}
\newtheorem*{prop}{Proposition}
\usepackage{array}
\usepackage{url}
\usepackage{adjustbox,lipsum}
\usepackage{romannum}
\usepackage{siunitx}
\usepackage{gensymb}
\usepackage{caption}
\usepackage [english]{babel}
\usepackage [autostyle, english = american]{csquotes}
\MakeOuterQuote{"}
\usepackage{float}
\def\footnoterule{\kern-3\p
  \hrule \kern 2.6\p} 
\makeatother

\begin{document}

\title{The Impact of Power-Electronics-Based Load Dynamics on Large-disturbance Voltage Stability}

\author{\IEEEauthorblockN{Daijiafan Mao, Karun Potty and Jiankang Wang}
\IEEEauthorblockA{Department of Electrical and Computer Engineering\\The Ohio State University\\
Columbus, Ohio 43210\\
mao.156@osu.edu; potty.1@osu.edu; wang.6536@osu.edu}}

\maketitle

\begin{abstract}
This paper establishes a new context where the power-electronics-based (PE-based) load, represented by Plug-in Electric Vehicles, dominates the total load composition in power systems. The inherent fast dynamics of PE-based load make conventional approaches of voltage stability analysis unsuitable. Under the new context, the mechanism and impacts of voltage instability under large disturbances have been analytically revealed. The Region of Attraction (ROA) of the stable equilibrium point has been estimated through nonlinear dynamical system theories, which implies a critical clearing time post grid disturbance. 
\end{abstract}

\section{Introduction}
Driven by power industry deregulation and evolution of electricity markets, the power grid has been pushed toward its operation limits of generation and transmission capability over the past two decades. As a result, power system voltage instability becomes a primary concern among utilities.

Voltage instabilities have been well studied in literature with both static and dynamic approaches \cite{yorino1992investigation,morison1993voltage,gao1996towards}. The static approach is based on power flow equations, estimating if the system is stable, and the stability margin by using voltage-reactive power sensitivity, eigenvalues of the power flow Jacobian and continuation power flow (CPF) techniques \cite{yorino1992investigation,sauer1990power}. The dynamic approach considers system dynamics induced by physical devices, such as generators, induction motors and self-restoring loads. The analysis techniques under dynamic approaches include time-domain simulation, linear/nonlinear system theories, etc. \cite{liu1989analysis}. It should be noted that the term "static" is inspired by the steady-state analysis tool, i.e., power flow. Nevertheless, the voltage stability analysis is dynamic by nature, as the driving force of voltage instability is the load's dynamical attempt to restore power consumption.

Hence, thorough studies have been conducted regarding the load-driven mechanism of voltage instability, either in generic forms of nonlinear models with first-order dynamics of voltage dependent loads, or in specific forms derived for individual devices (e.g., induction motor)\cite{xu1994voltage,van1998voltage}. The application of singular perturbation theory made possible the decomposition of entire system dynamics into two widely separate time scales \cite{kokotovic1999singular}, as shown in (\ref{perturb1}) and (\ref{perturb2}), enabling great simplification when corresponding analysis techniques have been applied.
\begin{align}
    \label{perturb1} \dot{x_S}&=f_S(x_S,x_F)\\
    \label{perturb2} \varepsilon \dot{x_F}&=f_F(x_S,x_F),
\end{align}
where, $x_S$, $x_F$ are vectors of slow and fast variables, respectively, and the system is separated into a slow subsystem (\ref{perturb1}) and fast subsystem (\ref{perturb2}) due to small parameter $\varepsilon$.

It has been shown that the transient response for the impedance network is very fast compared to the slower dynamics of load-driven voltage instability, and thus can be formally eliminated by setting $\varepsilon=0$. This approximation has reformed dynamical properties of the system, as (\ref{perturb2}) degenerates into the algebraic equation (\ref{algebraic}), which is also the equilibrium condition for the fast subsystem (\ref{perturb2}).
\begin{align}\label{algebraic}
    0=f_F(x_S,x_F)
\end{align}

Equations (\ref{perturb1}) and (\ref{algebraic}) constitute the Differential-Algebraic (D-A) system. In particular, for a long-term characteristic component such as a load tap changer (LTC), the time-scale decomposition simplifies the analysis through the quasi-steady-state (QSS) approximation by assuming all short-term dynamics, including synchronous generator, induction motor, etc., operate at equilibrium. This singularly perturbed model is shown in (\ref{decouple1}) - (\ref{decouple3}) \cite{van1998voltage}.
\begin{align}
    \label{decouple1}&\dot{z_c}=h_c(x,y,z_c)\\
    \label{decouple2}&0=f(x,y,z_c)\\
    \label{decouple3}&0=g(x,y,z_c),
\end{align}
where, without loss of generality, (\ref{decouple1}) models the long-term dynamics of continuous state variables $z_c$, (\ref{decouple2}) refers to the equilibrium condition of short-term dynamical states $x$ through singular perturbation, and (\ref{decouple3}) denotes the steady-state condition of network (i.e., power flow) for algebraic states $y$.

\begin{figure}[h]
\vspace{0em}
\centering
\includegraphics[scale=0.5]{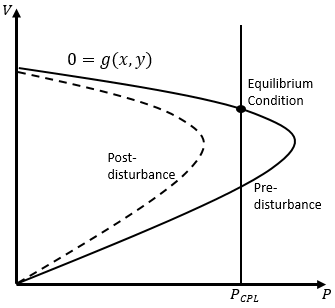}
\caption{$P-V$ Curve Demonstration of Voltage Instability. $P_{CPL}$ represents the static characteristic of constant power loads.}
\label{conventional}
\end{figure}

The fundamental assumption for this time-scale decomposition is that when large disturbances (e.g., loss of a transmission line or generation) are imposed, the response of electromagnetic transients on the impedance network always settle fast enough and are stable. Thus, before a disturbance, the grid operation point can be found by intersecting the Constant Power Load (CPL) characteristic curve with the network $P-V$ characteristics, as shown in Fig. \ref{conventional}. Post disturbance, the network characteristic shrinks drastically due to the change of network topologies. Thus, the CPL curve no long intersects the post-disturbance network $P-V$ curve, indicating voltage instability \cite{van1998voltage}.
\setlength\belowcaptionskip{1ex}

However, as an ongoing trend, the landscape of power system loads has been significantly altered by the power-electronics-based (PE-based) load, in particular, plug-in electric vehicles (PEV), gradually dominating the total load composition. According to the International Energy Agency (IEA), over 750 thousand fleets of new PEVs were registered in 2016 alone, and the worldwide PEV penetration target is 30\% of total market share by 2030\cite{global2017outlook}. The inherent high-bandwidth control of PE-based loads makes the assumption, on which the singular perturbation analysis is based, no longer valid. The dynamics of PE-based loads belong to neither long-term nor short-term characteristics with respect to conventional time-scale decoupling, but fall into one that is comparable to a network's instantaneous response. Therefore, we must resort to device-level analysis to analyze grid response under large grid disturbance.

As our first step toward studying the voltage stability under this new context, this paper investigates the interaction between a large grid-side disturbance and PEV load, which is modeled with a representative PE structure. The contributions of the paper are twofold. First, it demonstrates voltage instabilty induced from PE-based load response to grid-side disturbance. Second, it proposes a technique for estimating the Region of Attraction (ROA) of stable equilibria, which implies a critical clearing time for grid disturbance \cite{khalil1996noninear}. Despite exemplification with a rudimentary system, the proposed technique can be extended to analyze power grids of a greater scale. A time-domain simulation is presented in the last section.
\section{Modelling of PEV Charger and Dynamical Stability Analysis}
\subsection{System Configuration}
\begin{figure}[h]
\centering
\includegraphics[scale=0.5]{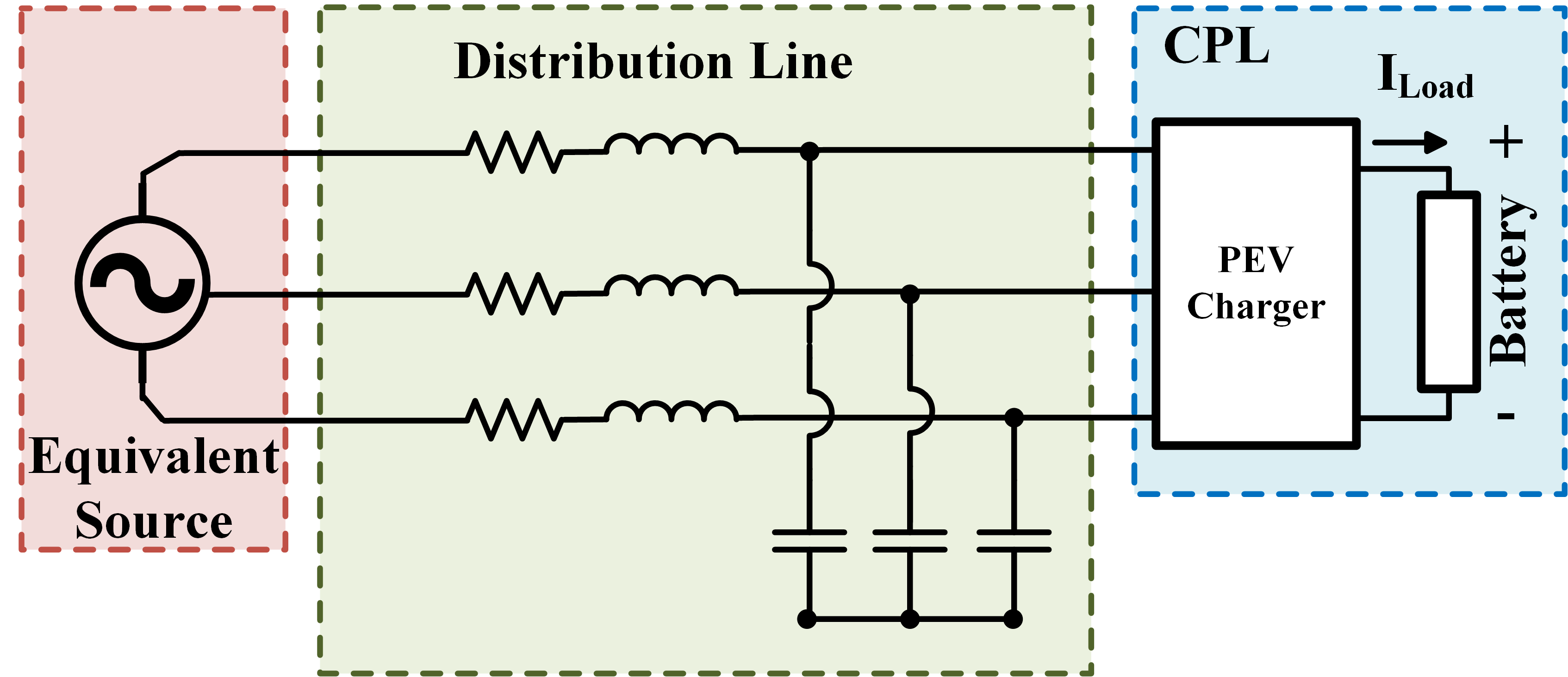}
\caption{System Configuration}
\label{configuration}
\end{figure}

Appropriate dynamical modeling of PEV loads is critical to study the voltage instability mechanism induced by large disturbances and to design preventive actions at load level. The individual PEV load is generally comprised of a cascaded system, which contains a grid-connected converter (AC/DC Rectifier), a load-side converter (DC/DC Converter) and a battery \cite{dharmakeerthi2015pev}. Such integrated configuration, as a whole, is designed to have constant power consumption at the load side regardless of input variations. Moreover, despite the reactive compensation requirement in certain applications, the high bandwidth PE controller generally ensures pure active power absorption from the grid with unity power factor. The general system configuration is shown in Fig. \ref{configuration}.
\subsection{Dynamical System Modelling} \label{mention}
\begin{figure}[ht]
\centering
\includegraphics[scale=0.5]{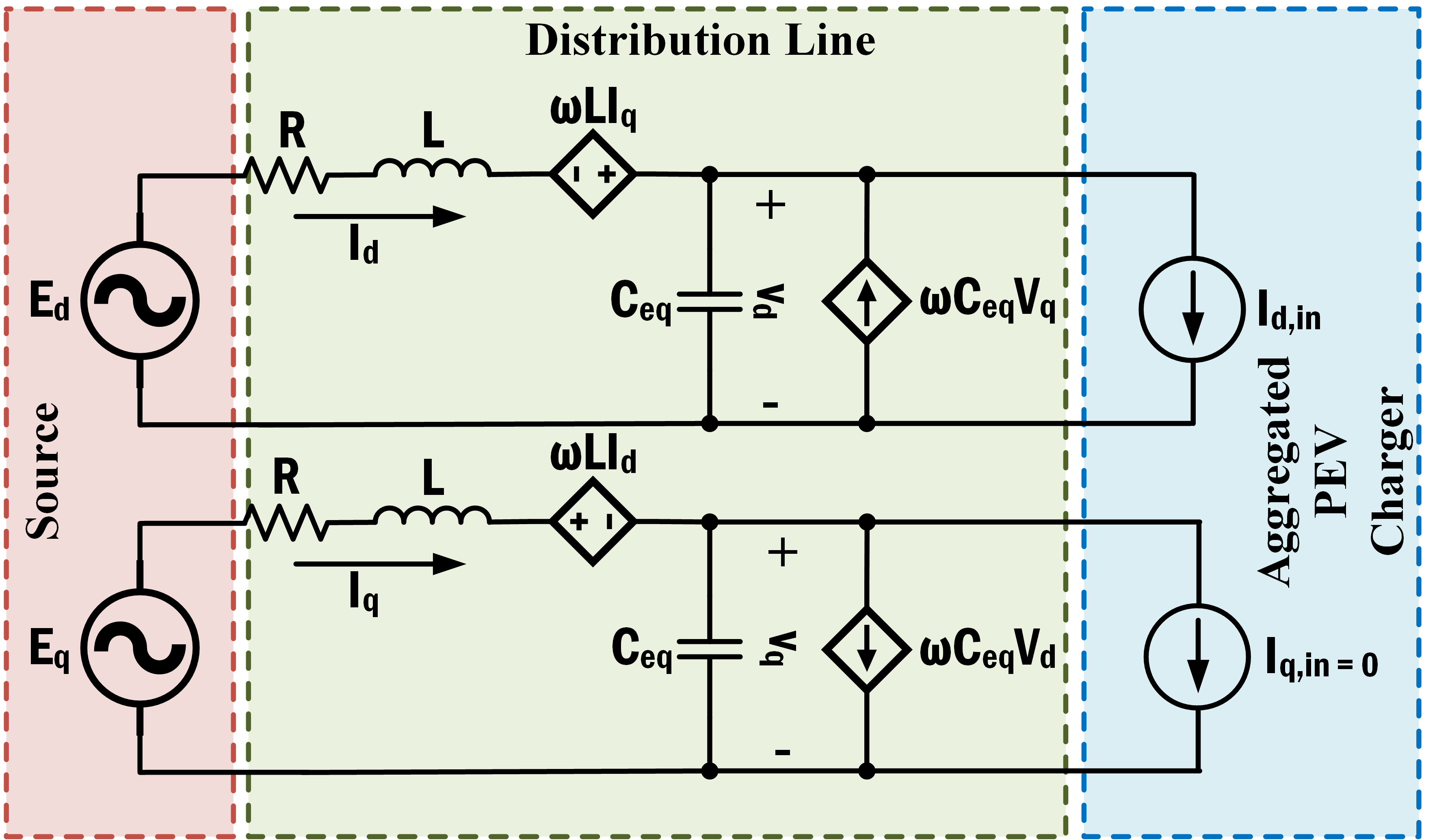}
\caption{System Equivalent Circuit in dq Frame}
\label{equivalent}
\end{figure}

The CPL characteristics are embedded with AC three-phase grid-connected system. A baseline model has been adopted, combining source and line dynamics with an infinite bandwidth CPL in order to simulate a PEV charger's short-term characteristics \cite{hiti1995modeling}. The system equivalent circuit in dq frame is shown in Fig. \ref{equivalent}.

The abc-to-dq transformation has been widely used in stability analysis of PE-based interface \cite{bazargan2014stability}. It makes all phase quantities fixed with respect to each other, thereby leading to constant self and mutual inductance. Thus, it reduces the complexity of AC grid-connected load system analysis and allows the whole system to be analyzed in a DC manner. The three-phase power consumption of a PEV charger is expressed in (\ref{power}). For convenience, we assume no q-axis current going through charger, $I_{q,in}=0$ , i.e., the charger is regulated at unity power factor. This assumption does not affect the conclusions of the analysis, and can be relaxed in the augmented form of \eqref{jacob}.
\begin{align}\label{power}
    P_{PEV}=\frac{3}{2}V_d I_{d,in}
\end{align}

Under the equivalent circuit, the system state model of an AC grid-connected PEV charger is
\begin{align}\label{system}
    \dot{\bold{x}}=f(\bold{x},\bold{u},\bold{p}),    
\end{align}
where $f:D \mapsto \mathbb{R}^4$ is continuously differentiable and $D\subset \mathbb{R}^4$ is a domain that contains the equilibrium point of (\ref{system}) $\bold{x}^* \in \mathscr{E} := \{\bold{x}^*\in D: f(\bold{x}^*)=0\}$.

By KVL and KCL, the detailed model is given in (\ref{eq1})-(\ref{eq4}). 
\begin{align}
    \label{eq1} &\dot{I_d}=-\frac{R}{L}I_d + \omega I_q-\frac{1}{L}V_d+\frac{E_d}{L}\\ 
    \label{eq2} &\dot{I_q}=-\omega I_d-\frac{R}{L}I_q-\frac{1}{L}V_q +\frac{E_q}{L}\\
    \label{eq3} &\dot{V_d}=\frac{1}{C_{eq}}I_d-\frac{2P_{PEV}}{3C_{eq}V_d}+\omega V_q\\
    \label{eq4} &\dot{V_q}=\frac{1}{C_{eq}}I_q-\omega V_d,
\end{align}
where the system state variables are line currents and load voltages $\bold{x}=[I_d ~I_q ~V_d ~V_q]^T$, the input variables $\bold{u}=[E_d ~E_q ~P_{PEV}]^T$ are grid (source) voltage and charger demand, and the system parameter set is $\bold{p} = \{\omega, ~R,~L,~C_{eq}\}$ where $\omega$ is the line frequency, $R,L$ are line resistance and inductance, respectively, and $C_{eq}$ is the equivalent capacitance representing combined shunt capacitance and charger's dynamical effects. 

The state Jacobian matrix of this highly-coupled, high-dimensional dynamical system can be expressed in (\ref{jacob}).
\begin{align}\label{jacob}
J=\frac{\partial f}{\partial \bold{x}}=
\begin{bmatrix}
-\frac{R}{L} & \omega & -\frac{1}{L} & 0\\
-\omega & -\frac{R}{L} & 0 & -\frac{1}{L}\\
\frac{1}{C_{eq}} & 0 & \frac{2P_{PEV}}{3C_{eq}V_d^2} & \omega\\
0 & \frac{1}{C_{eq}} & -\omega & 0
\end{bmatrix}
\end{align}

The distribution systems are characterized by their high $R/X$ ratio \cite{eminoglu2005new}. For instance, for a $12$ kV distribution line the ratio could be larger than $10$. Under this circumstance, we can approximate the reactance $X=\omega L \approx 0$ due to the line damping effect $\tau = R/L=R/(X/\omega)=R\omega /X>10 \omega$. Hence, the hereafter analysis is based on the decoupling assumption. Moreover, since the charger is operating in unity power factor (i.e., d-axis dominant), we assume that d-axis states ($V_d$ and $I_d$) are the main states, which dictate the major dynamics of the system, whereas q-axis states ($V_q$ and $I_q$) are independent elements that are coupled to each other but decoupled from active elements. Through this decoupling, the dynamical analysis can be conducted in the planar system $g:D\mapsto \mathbb{R}^2$ and the reduced model can be rendered as
\begin{align}
    &\dot{V_d}=-\frac{2P_{PEV}}{3C_{eq}V_d}+\frac{1}{C_{eq}}I_d\\
    &\dot{I_d}=-\frac{1}{L}V_d-\frac{R}{L}I_d +\frac{E_d}{L},
\end{align}
where the state variable set becomes $\bold{x}=[V_d ~I_d]^T$, input variables $\bold{u}=[E_d ~P_{PEV}]^T$, and $\bold{p} = \{R,~L,~C_{eq}\}$.

The corresponding reduced Jacobian matrix is expressed as
\begin{align}\label{reduce}
J_{r}=
\begin{bmatrix}
\frac{2P_{PEV}}{3C_{eq}V_d^2} & \frac{1}{C_{eq}}\\
-\frac{1}{L} & -\frac{R}{L}
\end{bmatrix}.
\end{align}
\subsection{Dynamical System Analysis}
The static stability of the operating condition can be obtained by observing the spectrum $spec(J_r)$ of (\ref{reduce}) evaluated at a specific parameter set $\bold{p}^*$ and equilibrium point $\bold{x}^* \in \mathscr{E} := \{\bold{x}^*\in D: g(\bold{x}^*)=0\}$. The sufficient condition for a particular loading value $P_{PEV}$ to be stable is such that $spec(J)\subset \mathbb{C}^-:=\{\lambda \in spec(J)\cap \mathbb{C}:~\Re [\lambda]<0 \}$. Hence, it is possible to find a demand upper bound $P_{PEV}^{max}$ and to identify the closeness of the current operating point to the voltage collapse point \cite{herrera2015stability}.

However, the above linearization-based approach is limited to analyzing the local stability at the equilibrium point. It can neither assess the robustness of equilibrium, nor predict the system behavior when (\ref{jacob}) becomes ill-conditioned. To understand the finite stability of the system, we have the following proposition.

\begin{figure*}[t]
\begin{subfigure}{0.33\textwidth}
\includegraphics[width=1\linewidth, height=4cm]{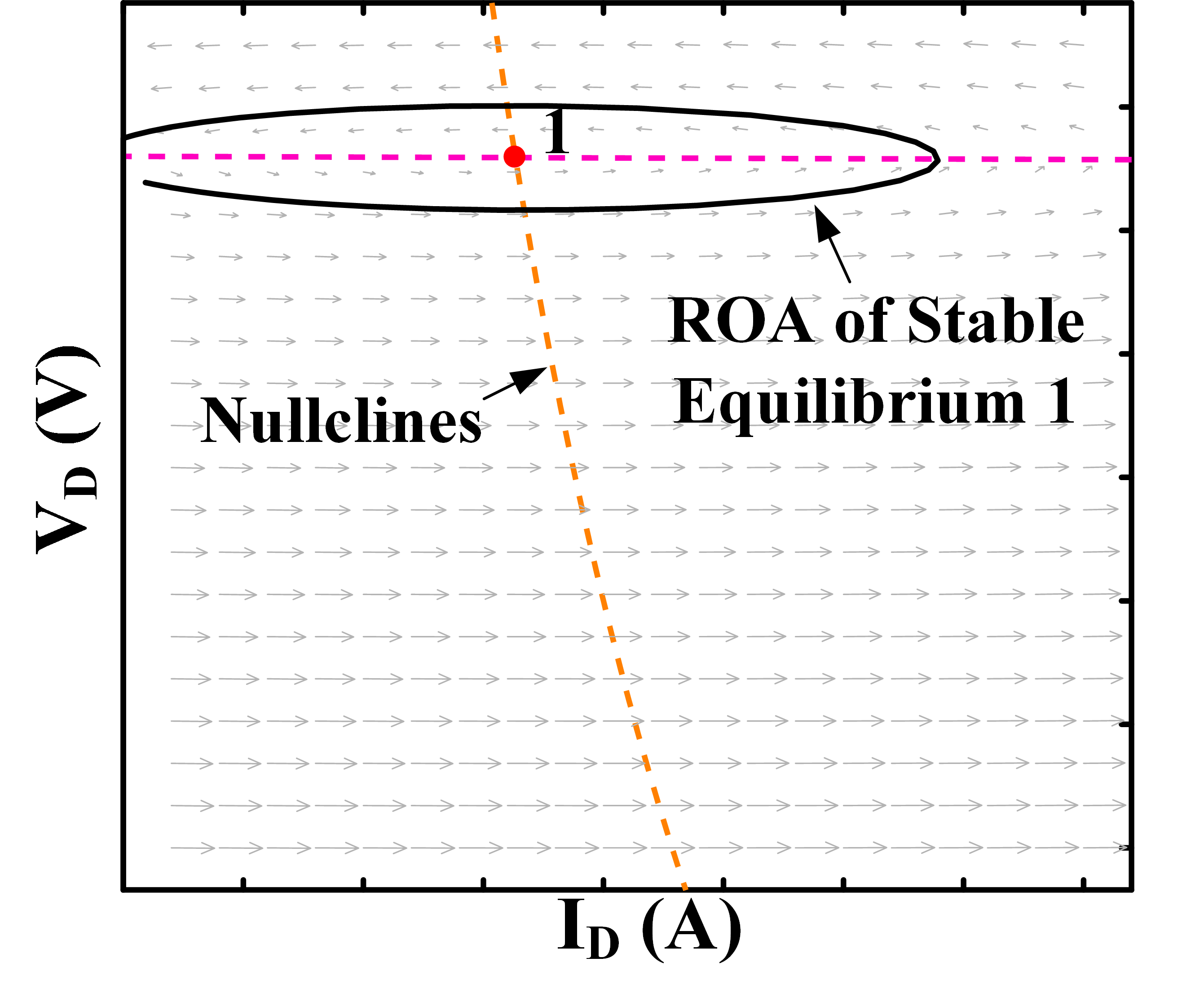} 
\caption{Initial Stable Operation}
\label{scheme1}
\end{subfigure}
\begin{subfigure}{0.33\textwidth}
\includegraphics[width=1\linewidth, height=4cm]{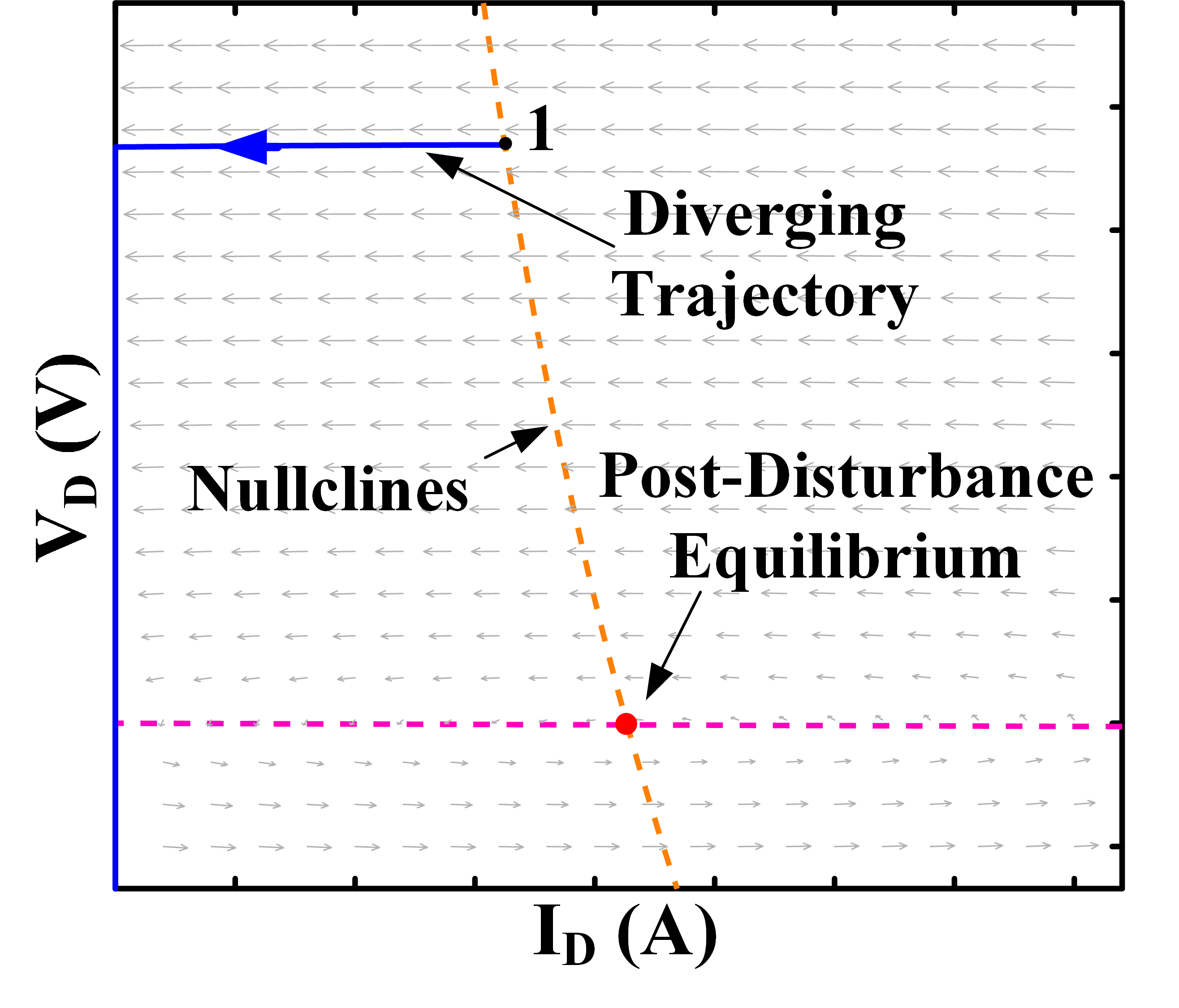}
\caption{Post-Disturbance Unstable Operation}
\label{scheme2}
\end{subfigure}
 \begin{subfigure}{0.33\textwidth}
\includegraphics[width=1\linewidth, height=4cm]{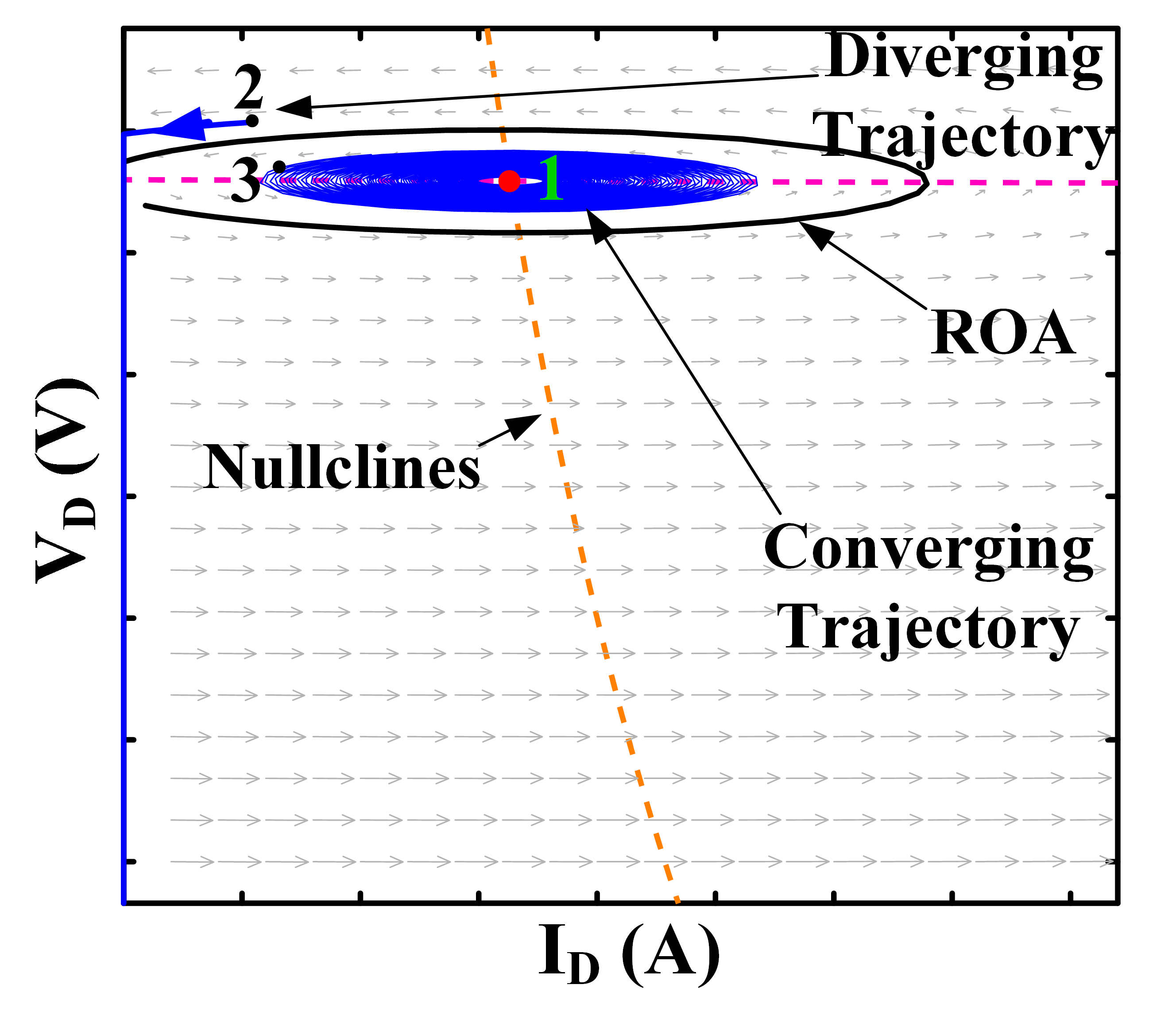} 
\caption{Trajectory Interaction with ROA}
\label{scheme3}
\end{subfigure}
\caption{Phase Portrait Analysis}
\label{scheme}
\end{figure*}

\begin{prop}
The structure of three-phase charger dynamics determines that the system exhibits unstable limit cycle $\Gamma (\bold{x}_0)=\{\bold{x}\in D: ~\bold{x}=\Phi _t^g (\bold{x}_0), ~t\in [0,+\infty]\}$, where $\Phi _t^g (\bold{x}_0)$ is the flow of the state space model $\dot{\bold{x}}=g(\bold{x},\bold{u},\bold{p})$ passing through initial condition $\bold{x}_0$. The interior of this limit cycle implies a Region of Attraction (ROA) $\mathcal{A}$ of a Locally Asymptotically Stable (LAS) equilibrium, where $\mathcal{A}:=\textbf{int}(\Gamma (\bold{x}_0))=\{\bold{x}_0 \in \mathbb{R}^2 :~\lim \limits_{t\to \infty} |\bold{x}(t,\bold{x}_0)|=\bold{x}^* \},~\forall \bold{x}^* \in \mathscr{E} ~s.t. ~\bold{x}^* \text{ is } LAS$.
\end{prop}

The ROA $\mathcal{A}$ of an initial stable equilibrium is an open, connected and invariant set bounded by unstable limit cycle, as shown in the illustrative phase portrait Fig. \ref{scheme1}, where for any initial operating condition $(I_{d0},V_{d0})$ starting inside the ROA, the system trajectory will spiral toward the stable equilibrium point $1$. Suppose the system undergoes a large disturbance that makes the post-disturbance equilibrium unstable, then the original stable equilibrium point $1$ becomes the initial condition for the new unstable system. This results in the trajectory devolving away, as shown in Fig. \ref{scheme2}. When the system is restored to original stable operation through fault-clearing, whether or not the trajectory converges depends on where the system operating point is when fault-clearing occurs. For example, as shown in Fig. \ref{scheme3}, if the fault is cleared when diverging state trajectory reached point $2$ (outside ROA), then it diverges away; on the other hand, if the fault is cleared at point $3$ (within ROA), the state trajectory will converge back to stable equilibrium $1$ and the fault-clearing is successful.

\begin{figure}[h!]
\begin{subfigure}{0.5\textwidth}
\includegraphics[width=0.95\columnwidth,height=5.5cm]{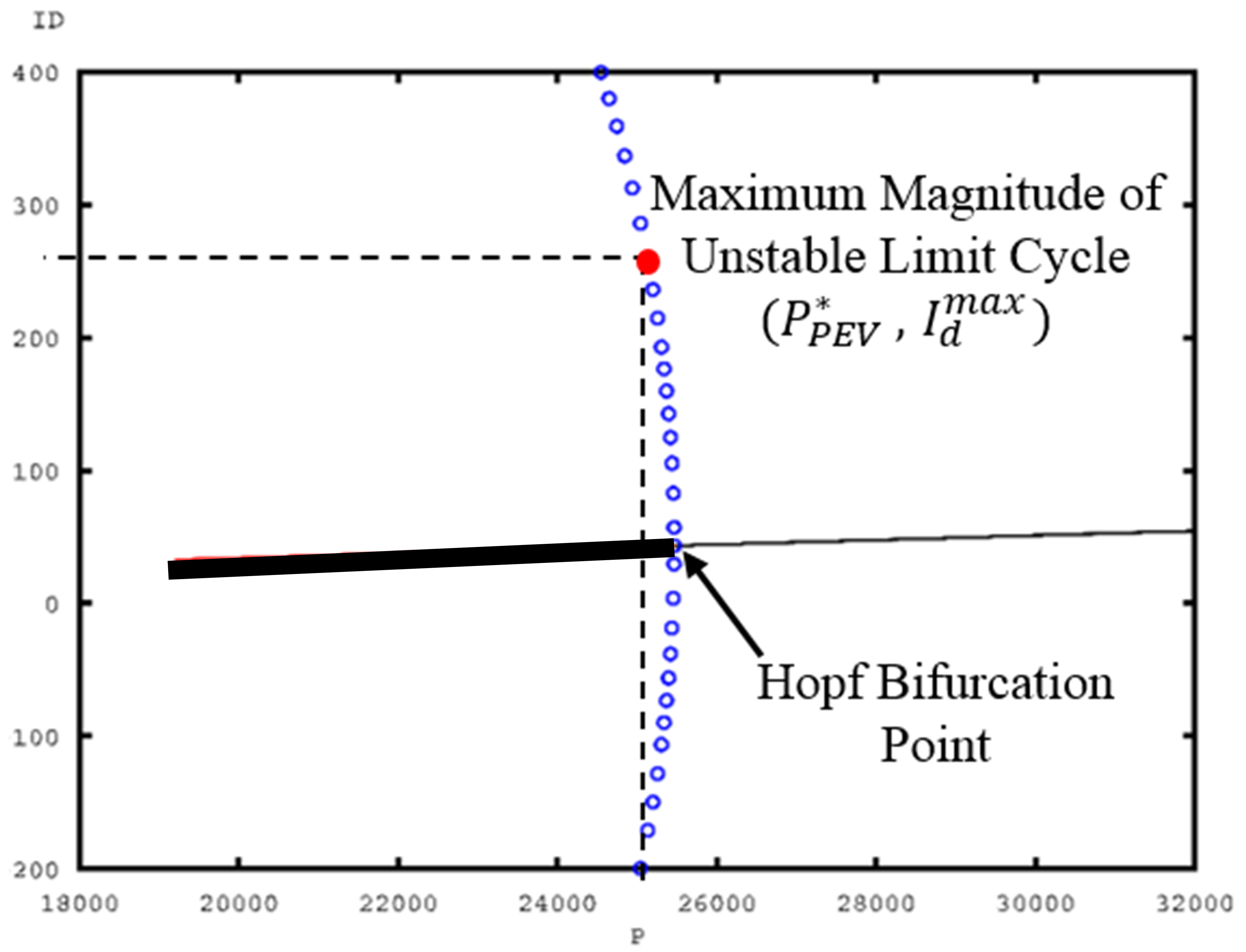} 
\caption{$I_d$ \text{ vs. } $P_{PEV}$}
\label{bdi}
\end{subfigure}
\begin{subfigure}{0.5\textwidth}
\includegraphics[width=0.95\columnwidth,height=5.5cm]{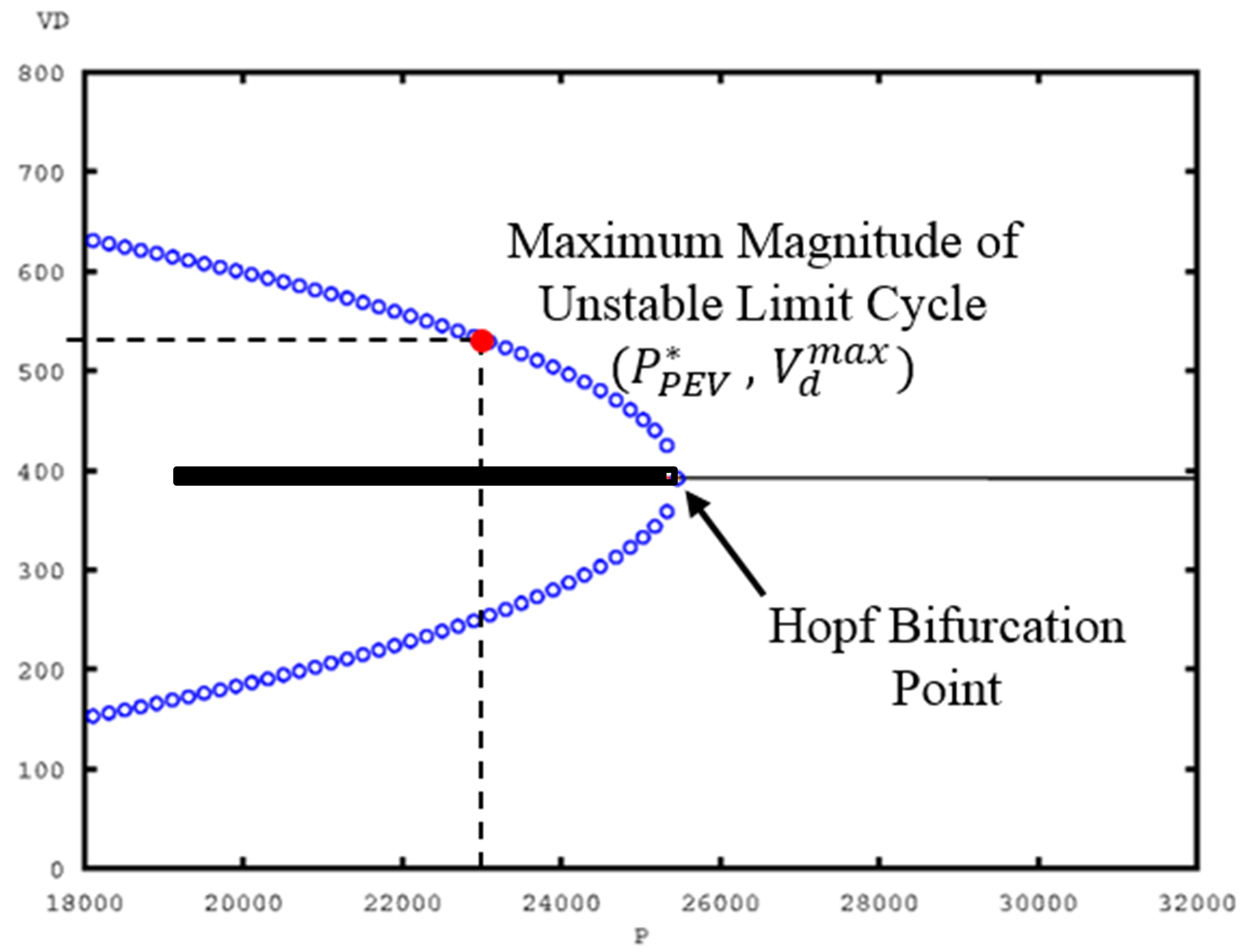}
\caption{$V_d$ \text{ vs. } $P_{PEV}$}
\label{bdv}
\end{subfigure}
\caption{Bifurcation Diagram of Parameter $P_{PEV}$}
\label{bif}
\end{figure}

The aforementioned proposition is confirmed through bifurcation diagram analysis as shown in Fig. \ref{bif}, obtained by simulating a system with the following parameters also used throughout Section \Romannum{3}: $R=0.0064 ~\Omega$, $L=1.698 ~\mu H$, $C_{eq}=29.333 ~\mu F$, $E_d=392.125 ~V$, $P_{PEV}=19200 ~W$. These values were selected to make the initial operating condition start at stable (LAS) equilibrium and be realistically consistent with short-line distribution-connected PEV charger model described in Section \ref{mention}.

In the bifurcation diagram, as parameter $P_{PEV}$ varies, the stable state ($I_d$ or $V_d$) branch is denoted by the thick solid curve, and an unstable branch by the thin curve. Changes in stability occurs at bifurcation point, where the stability of an equilibrium is lost through its interaction with a limit cycle. The unstable limit cycle exists prior to bifurcation, shrinks and eventually disappears as it coalesces with a stable equilibrium at bifurcation point. Afterwards, the equilibrium becomes unstable, resulting in growing oscillating instability. This series of behavior falls in the case of subcritical Hopf Bifurcation (HB). The circles emanating from the Hopf bifurcation point yield an estimation of maximum magnitude of unstable limit cycle under a specific value of $P_{PEV}$.

The validation of HB is reflected in the reduced state Jacobian $J_r$ as in (\ref{reduce}). In particular, the HB occurs when a pair of complex conjugate eigenvalues lies exactly on the imaginary axis. However, the HB analysis provides more information than standard linear stability theory because it also takes account of the effect of non-linearity and predicts oscillatory instability through limit cycle interaction. Therefore, such bifurcation analysis is best thought of as a supplementary tool that helps to explain the form of the instability when the stable operating equilibrium is lost.

\section{Case Study}
In time domain analysis, the ROA implies a critical fault clearing time, after which the state trajectory originating from the pre-disturbance stable equilibrium have already drifted out of ROA and are not able to converge. Note that the ROA and trajectories are assumed to be bounded in first quadrant $(I_d,V_d)\subset \mathbb{R}_+$ based on the realistic case that the power flow is unidirectional, i.e., from source to PEV load. 

The occurrence and clearance of non-small grid disturbance could be a tripping and ensuing re-closing of the generation/transmission equipment or sudden incremental aggregation of consumption. In particular, two scenarios have been considered as non-small disturbance in this paper: (i) sudden source voltage drop, which means a decrease of input variables $E_d$ and $E_q$; (ii) power demand surge, which means an increase of input variable $P_{PEV}$. The system response under disturbance has been simulated through MATLAB.

\begin{figure}[h]
\begin{subfigure}{0.5\textwidth}
\includegraphics[width=0.95\columnwidth, height=4.2cm]{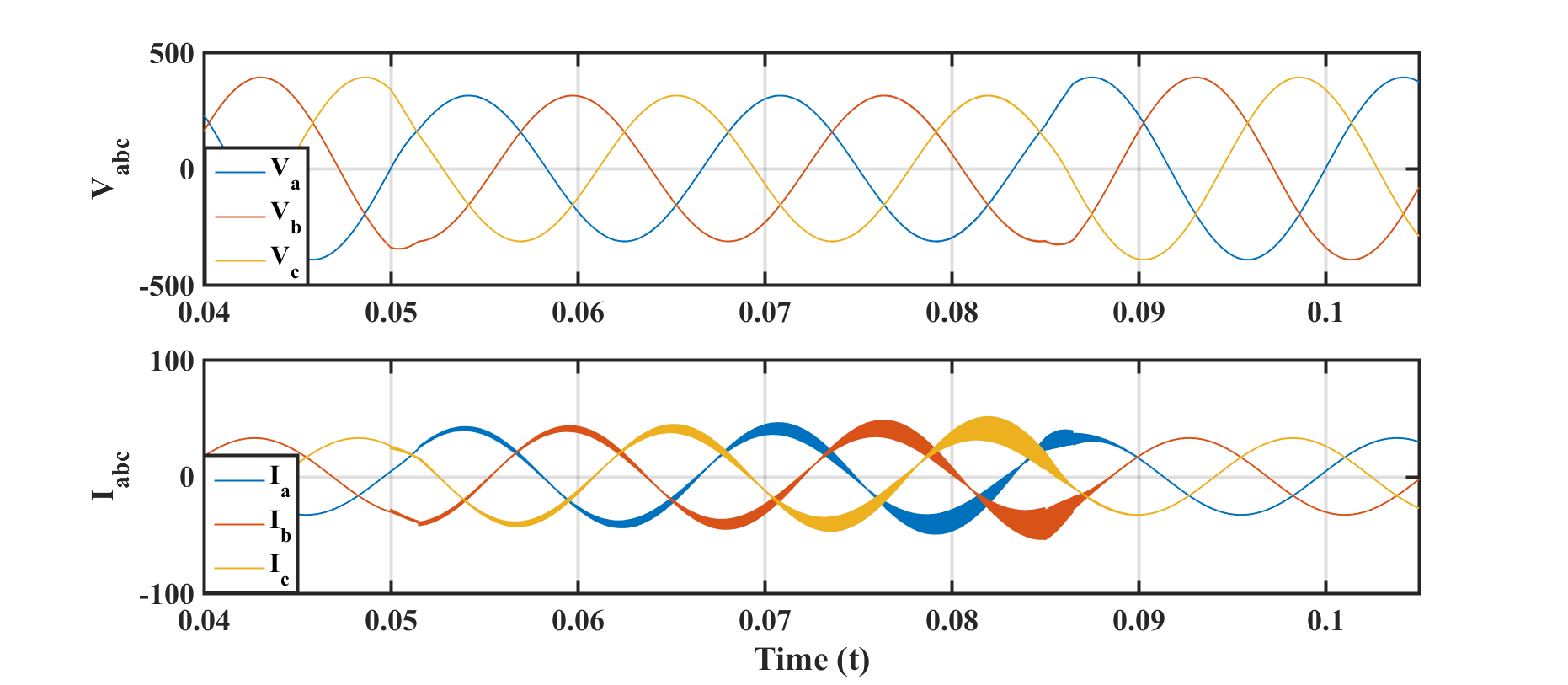} 
\caption{Stable Response when $t_{clear}=0.085 ~s$}
\label{stable}
\end{subfigure}
\begin{subfigure}{0.5\textwidth}
\includegraphics[width=0.95\columnwidth, height=4.2cm]{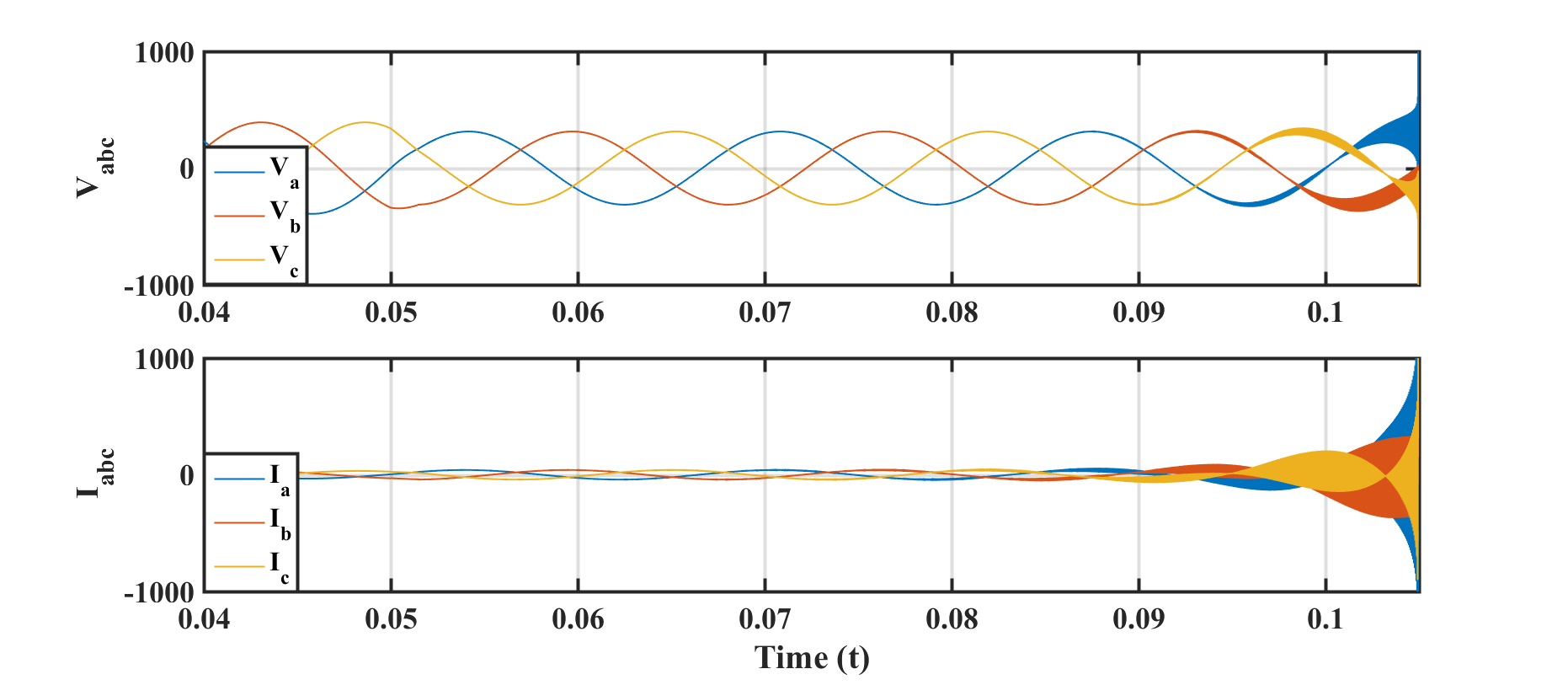}
\caption{Unstable Response when $t_{clear}=0.15 ~s$}
\label{unstable}
\end{subfigure}
\caption{Critical Clearing Time Implication for Scenario (i)}
\label{simV}
\end{figure}

\begin{figure}[h]
\begin{subfigure}{0.5\textwidth}
\includegraphics[width=0.95\columnwidth, height=4.2cm]{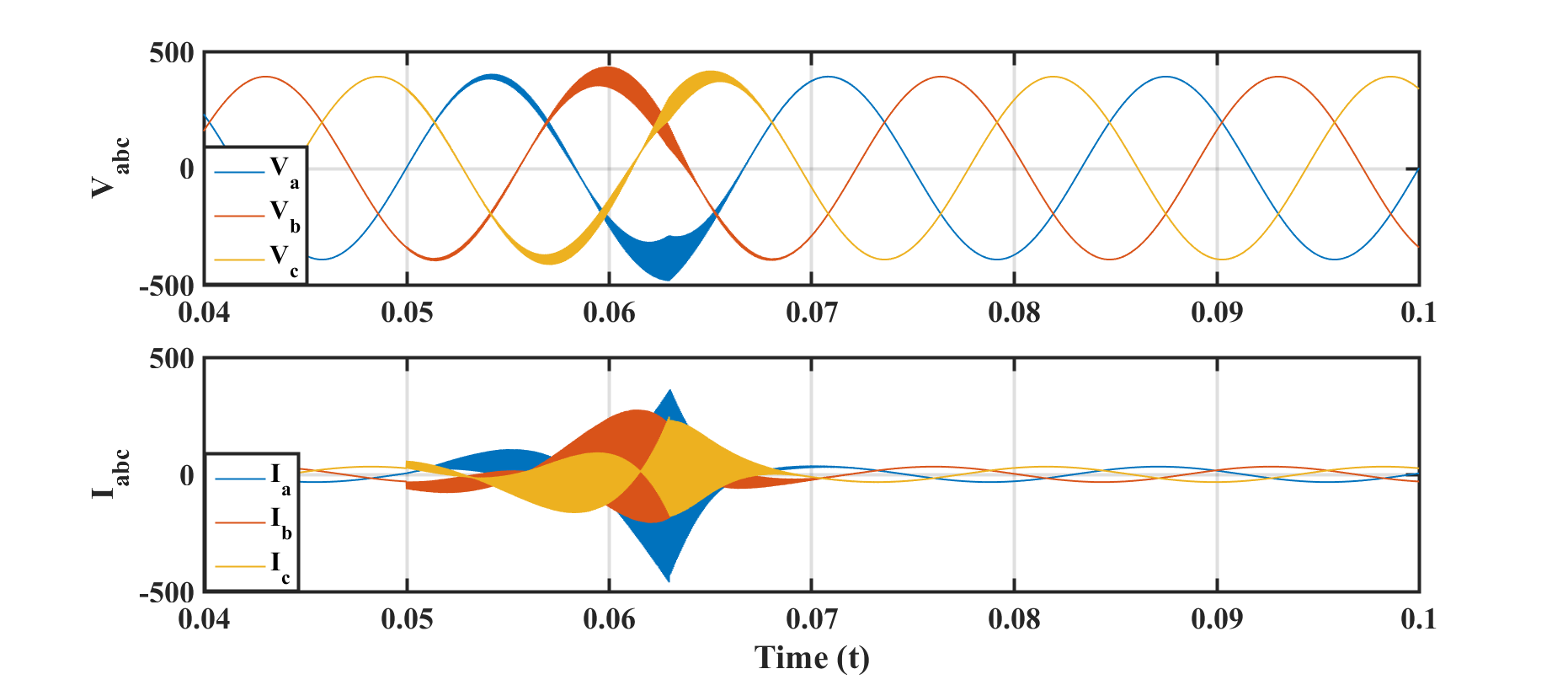} 
\caption{Stable Response when $t_{clear}=0.068 ~s$}
\label{stable2}
\end{subfigure}
\begin{subfigure}{0.5\textwidth}
\includegraphics[width=0.95\columnwidth, height=4.2cm]{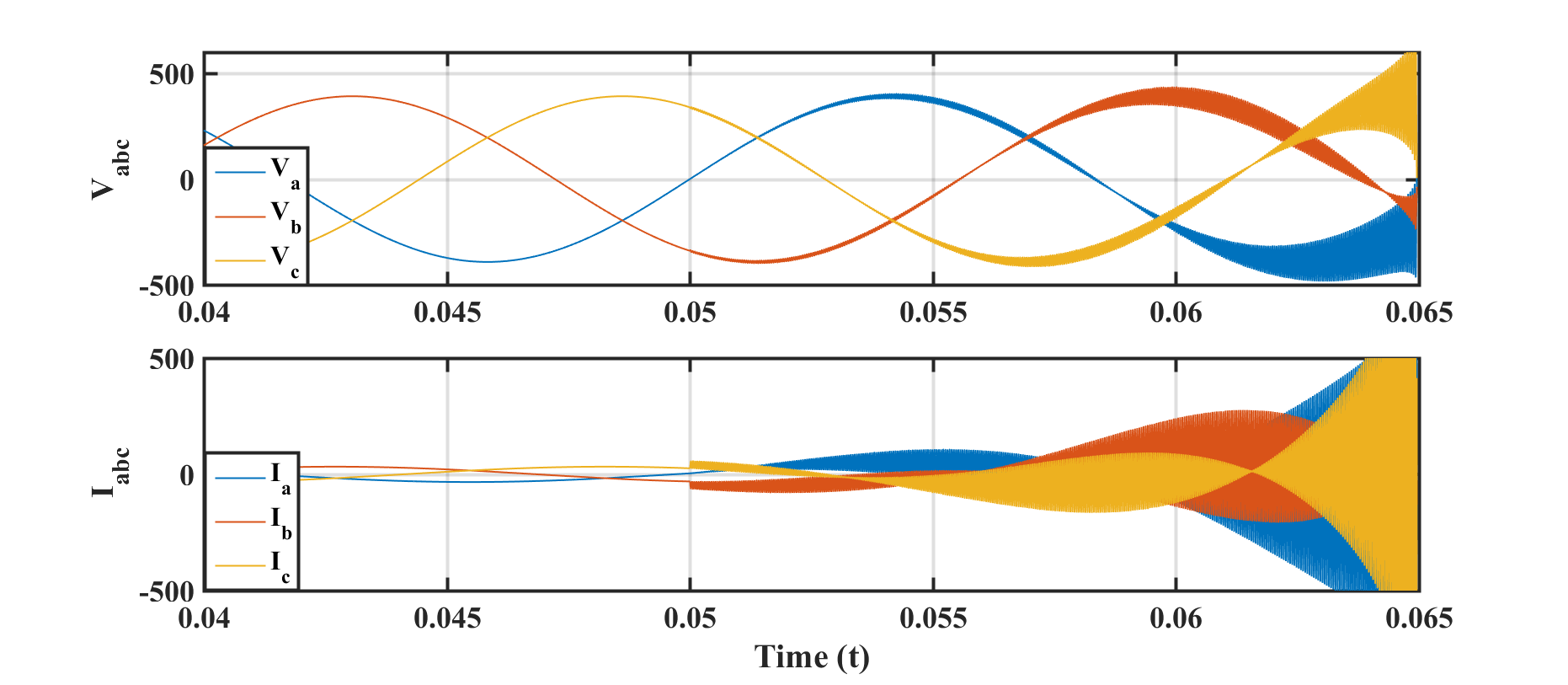}
\caption{Unstable Response when $t_{clear}=0.08 ~s$}
\label{unstable2}
\end{subfigure}
\caption{Critical Clearing Time Implication for Scenario (ii)}
\label{simP}
\end{figure}

For both scenarios, the disturbance occurs at $t=0.05 ~s$ and is cleared at $t_{clear}$. The three-phase line voltage and current response for scenarios (i) and (ii) have been observed in Fig. \ref{simV} and Fig. \ref{simP}, respectively. In scenario (i), the fault clearing times are selected at $t_{clear}=0.085 ~s$ and $t_{clear}=0.15 ~s$, respectively. As observed in Fig. \ref{stable}, both voltage and current resume pre-disturbance values after a certain tolerable perturbation. However, when the fault clearing has been delayed to $t_{clear}=0.15 ~s$, the line state will exceed the acceptable range, as shown in Fig. \ref{unstable}, which will trip the protection in the charger. Such dynamical loss of stability implies that the critical clearing time for scenario (i) is within the range $t_{cr}^A\in (0.085,0.15)s$. Similar implication has been observed for scenario (ii), with $t_{cr}^B\in (0.068,0.08)s$. 

Estimation of the critical clearing time is more complicated and will be detailed in future works. A general direction is to integrate searching algorithms with the Boundary Controlling Unstable (BCU) Equilibrium Point method \cite{chiang1994bcu}. The system's energy functions can be constructed at the closest and farthest unstable equilibria, respectively, corresponding to the inner and outer perimeters of the ROA. A searching algorithm, e.g., bisection method, can be applied with the starting points defined as the spheres of the two energy functions to estimate the flee time (critical clearing time).

\section{Conclusion}
The inherent fast dynamics of PE-based loads make conventional approaches of voltage stability analysis unsuitable. This paper is our first step toward understanding the voltage instability mechanisms in future power systems, wherein PE-based loads dominate the total power consumption. By studying a PEV-connected rudimentary system, we analyze the mechanism and impact of dynamical loss of voltage stability under grid-side disturbances. The Region of Attraction (ROA) of the stable equilibrium condition is estimated through nonlinear dynamical system theories, which implies a critical clearing time for grid disturbance. This implication is demonstrated through time-domain simulation.

\bibliographystyle{IEEEtran}
\bibliography{ref/reference}


\end{document}